\documentclass{PoS}
\usepackage{graphicx}
\newcommand{\ep}{\varepsilon}

\title{Feynman Diagrams, Differential Reduction, and Hypergeometric Functions}

\ShortTitle{Feynman Diagrams, Differential Reduction, and Hypergeometric Functions}

\author{\speaker{Mikhail Kalmykov}%
        \thanks{Fellow of the Conference Organizing Committee.}\\
II. Institut f\"ur Theoretische Physik, Universit\"at Hamburg, 
Luruper Chaussee 149, 22761 Hamburg, Germany \\
Joint Institute for Nuclear Research,$141980$ Dubna (Moscow Region), Russia \\
E-mail: \email{kalmykov.mikhail@gmail.com}}

\author{Vladimir~V.~Bytev\thanks{Research supported by MK 1607.2008.2}\\
Joint Institute for Nuclear Research,$141980$ Dubna (Moscow Region), Russia \\
II. Institut f\"ur Theoretische Physik, Universit\"at Hamburg, 
Luruper Chaussee 149, 22761 Hamburg, Germany \\
E-mail: \email{bvv@mail.jinr.ru}}

\author{Bernd~A.~Kniehl \\
II. Institut f\"ur Theoretische Physik, Universit\"at Hamburg, 
Luruper Chaussee 149, 22761 Hamburg, Germany \\
E-mail: \email{kniehl@mail.desy.de}}

\author{B.F.L.~Ward 
\\
Department of Physics, Baylor University, One Bear Place, Waco, TX 76798, USA \\
E-mail: \email{BFL{\underline\ }Ward@baylor.edu}}

\author{Scott A.~Yost \\
Department of Physics, The Citadel, 171 Moultrie St., Charleston, SC 29409, USA \\
E-mail: \email{scott.yost@citadel.edu }}

\abstract{
We will present some (formal) arguments that any Feynman diagram 
can be understood as a particular case of a sum of Horn-type multivariable 
hypergeometric functions. 
The advantages and disadvantages of this type of approach 
to the evaluation of Feynman diagrams is discussed. 
}

\FullConference{XII Advanced Computing and Analysis Techniques in Physics Research\\
                November 3-7, 2008\\
                Erice, Italy}

\begin{document}
\section{Introduction}
Any dimensionally-regularized \cite{dimreg} multiloop Feynman diagram 
with propagators $1/(p^2-m^2)$
can be written in the form of a finite sum of multiple Mellin-Barnes integrals 
\cite{calan,MB} obtained via a Feynman-parameter or ``$\alpha$'' 
representation: \cite{QFT}
\begin{eqnarray}
F\left(a_{js},b_{km},c_i,d_j,\vec{x} \right) = 
\int_{\gamma+i\mathbb{R}} dz_1 \ldots dz_r
\frac{\Pi_{j=1}^{p} \Gamma\left(\sum_{s=1}^r a_{js} z_s+c_j\right)}
     {\Pi_{k=1}^{q} \Gamma\left(\sum_{m=1}^r b_{km} z_m+d_k\right)}
x_1^{-z_1} \ldots x_r^{-z_r} \; , 
\label{MB}
\end{eqnarray}
where 
$
a_{js}, b_{km}  \in \mathbb{Q}, c_i, d_j \in \mathbb{C},\ 
$
and $\gamma$ is chosen such that the integral exists. 
In this integral, the $x_a$ are rational functions of kinematic invariants 
(masses and  momenta) and the matrices $a_{js}, b_{kr}$ and 
vectors $c_i, d_j$ depend linearly on the dimension $n$ of 
space-time.\footnote{
Starting from the $\alpha$-representation for 
Feynman diagrams  with (irreducible) numerators 
\cite{QFT}, the Mellin-Barnes representation can be written 
explicitly in terms of Symanzik polynomials.\cite{calan}
In Refs.\ \cite{numerator:massless,numerator:massive}, it
was shown that this $\alpha$-representation can be understood as a
linear combination of scalar Feynman diagrams of the original type 
with shifted powers of propagators and space-time dimension multiplying a
tensor factor depending on external momenta.
Through fifty years of the evaluation of Feynman diagrams, many auxiliary 
programs have been created for the generation of Symanzik polynomials.
In particular, the matrix representation for these polynomials derived 
by Nakanishi (see Eqs.~(3.23)$-$(3.34) in Ref.\ \cite{Nakanishi}) 
has been realized on FORM by Oleg Tarasov in Ref.\ \cite{Symanzik1},
by Oleg Veretin in \cite{Symanzik2}, and recently in Ref.\ \cite{Symanzik3}.
}.  
In dimensional regularization, $n$ is treated as an arbitrary parameter, 
so that $n=I-2\ep$ for a positive integer $I$. 

Formally,\footnote{
This is true under the condition that there is a 
common domain of convergence. 
For particular values of the kinematic variables and 
powers of the propagators, the Mellin-Barnes representation may contain 
terms like $\hbox{$\Gamma(a-s)$} \times \hbox{$\Gamma(a+s)$}$ or 
$\Gamma^2(a-s) \times \Gamma^2(b+s)$, where $a$ and $b$ are parameters and 
$s$ is an integration variable. 
Two algorithms for the practical construction of the $\ep$-expansion 
of Mellin-Barnes integrals with such singularities
have been described in Refs.~\cite{Smirnov-Tausk},
These are now implemented in several packages \cite{Gzakon}.
It will be quite interesting to apply modern mathematical algorithms, 
as was done in Ref.\ \cite{hironaka}, 
to the problem of singularities in the Mellin-Barnes representation as well. 
In the framework of our approach, such singularities can be regularized 
by introducing an additional analytical regularization for each propagator 
(see Ref.\ \cite{extra}) or by introducing masses to regulate IR singularities. 
General theorems on the properties of Feynman diagrams in dimensional 
regularization guarantee that a smooth limit of the auxiliary regularization should exist.
} 
this integral can be expressed in terms of 
a sum of residues of the integrated expression 
\begin{eqnarray}
F\left(a_{js},b_{km},\vec{c},\vec{d},\vec{\alpha}, \vec{x} \right) = 
\sum_{\vec{\alpha}} B_{\vec{\alpha}} \vec{x}^{\;\vec{\alpha}} 
\Phi(\vec{\gamma};\vec{\sigma};\vec{x}) \;, 
\label{FI}
\end{eqnarray}
where the components of the vector $\vec{\alpha}$
are defined in terms of the matrix components $a_{js},b_{km}$ 
and vectors $\vec{c},\vec{d}$, the coefficients 
$
B_{\vec{\alpha}}
$
are ratios of $\Gamma$-functions with arguments depending on $\vec{\alpha}$, 
and the functions $\Phi(\vec{\gamma};\vec{\sigma};\vec{x})$ have
the form
\begin{eqnarray}
\Phi(\vec{\gamma};\vec{\sigma};\vec{x}) 
= 
\sum_{m_1,m_2,\cdots, m_r=0}^\infty 
\Biggl( 
\frac{
\Pi_{j=1}^K
\Gamma\left( \sum_{a=1}^r \mu_{ja}m_a+\gamma_j \right)
}
{
\Pi_{k=1}^L
\Gamma\left( \sum_{b=1}^r \nu_{kb}m_b+\sigma_k \right)
}
\Biggr) 
x_1^{m_1} \cdots x_r^{m_r} \;,
\label{Phi}
\end{eqnarray}
with
$
\mu_{ab}, \nu_{ab} \in \mathbb{Q},\ 
\gamma_j,\sigma_k \in \mathbb{C}. 
$

Let 
$
\vec{e}_j = (0,\cdots,0,1,0,\cdots,0)
$ 
denote the unit vector with unity in its $\j^{\rm th}$ entry, and define
$
\vec{x}^{\vec{m}} = x_1^{m_1} \cdots x_r^{m_r} 
$
for any integer multi-index 
$\vec{m} = (m_1, \cdots, m_r)$. 
In accordance with the definition of Refs.\ \cite{bateman,Gelfand}, 
the (formal) multiple series 
$
\sum_{\vec{m}=0}^\infty C(\vec{m}) \vec{x}^{\vec{m}},
$
is called {\it hypergeometric} if, for each $i=1, \cdots, r$, the ratio 
$
C(\vec{m}+\vec{e}_i)/C(\vec{m})$
is a rational function in the index of summation $(m_1, \cdots, m_r)$.
Ore and Sato \cite{Ore:Sato} found (see also Ref.\ \cite{Gelfand})
that the coefficients of such a series have the general form 
\begin{eqnarray}
C(\vec{m})
= \Pi_{i=1}^r \lambda_i^{m_i}R(\vec{m})
\Biggl( 
\frac{
\Pi_{j=1}^N \Gamma(\mu_j(\vec{m})+\gamma_j)
}
{
\Pi_{k=1}^M \Gamma(\nu_k(\vec{m})+\delta_k)
}
\Biggr) \;,
\label{ore}
\end{eqnarray}
where 
$
N, M \geq 0, 
$
$
\lambda_j,\delta_j, \gamma_j \in \mathbb{C} 
$
are arbitrary complex numbers, 
$\mu_j, \nu_k: \mathbb{Z}^r \to \mathbb{Z}$ are arbitrary integer-valued 
linear maps,  and $R$ is an arbitrary rational function. 
A series of this form is called a {\it Horn-type} hypergeometric series.

We deduce from Eqs.\ (\ref{Phi}) and (\ref{ore})
that, assuming there is a region of variables
where each multiple series of Eq.\ (\ref{FI}) is convergent, any
Feynman diagram can be understood as a special case of a Horn-type 
hypergeometric function with the functions $R(\vec{m})$ equal to 
unity.\footnote{Using the technique presented in Ref.\ \cite{Melnikov}, 
this statement can also be shown to be valid for the phase-space integral.}
One interesting property of Horn-type hypergeometric functions is the
existence of a set of differential contiguous relations between functions
with shifted arguments. We consider such relations in the following
section. 

\section{Contiguous Relations via Linear Differential Operators}
Let us consider a formal series of type (\ref{Phi}).
The sequences $\vec{\gamma}=(\gamma_1,\cdots, \gamma_K)$
and $\vec{\sigma}=(\sigma_1,\cdots, \sigma_L)$ are called {\it upper }
and {\it lower} parameters of the hypergeometric function, respectively. 
Two functions of type (\ref{Phi})
with sets of parameters shifted by a unit, 
$\Phi(\vec{\gamma}+\vec{e_c};\vec{\sigma};\vec{x})$ and 
$\Phi(\vec{\gamma};\vec{\sigma};\vec{x})$, 
are related by a linear differential operator:
\begin{eqnarray}
\Phi(\vec{\gamma}+\vec{e_c};\vec{\sigma};\vec{x}) 
& = & 
\sum_{m_1,m_2,\cdots, m_r=0}^\infty 
\Biggl( 
\frac{
(\sum_{a=1}^r \mu_{ca}m_a+\gamma_c)
\Pi_{j=1}^K
\Gamma\left( 
\sum_{c=1}^r \mu_{jc}m_c+\gamma_j \right)
}
{
\Pi_{k=1}^L
\Gamma\left( \sum_{b=1}^r \nu_{kb}m_b+\sigma_k \right)
}
\Biggr) 
x_1^{m_1} \cdots x_r^{m_r} 
\nonumber \\ 
& = &  
\sum_{m_1,m_2,\cdots, m_r=0}^\infty 
\Biggl( 
\frac{
\left ( \sum_{a=1}^r 
 \mu_{ca} x_a \frac{\partial}{\partial x_a}+\gamma_c \right)
\Pi_{j=1}^K
\Gamma\left ( \sum_{c=1}^r \mu_{jc}m_c+\gamma_j \right)
}
{
\Pi_{k=1}^L
\Gamma\left( \sum_{b=1}^r \nu_{kb}m_b+\sigma_k \right)
}
\Biggr) 
x_1^{m_1} \cdots x_r^{m_r} 
\nonumber \\ 
& = &  
\left ( \sum_{a=1}^r \mu_{ca} x_a \frac{\partial}{\partial x_a}+\gamma_c \right)
\Phi(\vec{\gamma};\vec{\sigma};\vec{x}) 
\equiv 
U_{[\gamma_c \to \gamma_c+1]}^+
\Phi(\vec{\gamma},\vec{\sigma}, \vec{x}) 
\;.
\label{do1}
\end{eqnarray}
Similar relations also exist for the lower parameters: 
\begin{eqnarray}
\Phi(\vec{\gamma};\vec{\sigma}-\vec{e}_c;\vec{x}) 
& = & 
\sum_{m_1,m_2,\cdots, m_r=0}^\infty 
\sum_{b=1}^r (\nu_{cb}m_b+\sigma_c)
\Biggl( 
\frac{
\Pi_{j=1}^K
\Gamma\left(\sum_{a=1}^N \mu_{ja}m_a+\gamma_j \right)
}
{
\Pi_{k=1}^L
\Gamma\left(\sum_{b=1}^r \nu_{kb}m_b+\sigma_k \right)
}
\Biggr) 
x_1^{m_1} \cdots x_r^{m_r} 
\nonumber \\ 
& = &  
\left( 
\sum_{b=1}^r 
\nu_{cb} x_b \frac{\partial}{\partial x_b} + \sigma_c
\right)
\Phi(\vec{\gamma};\vec{\sigma};\vec{x}) 
\equiv 
L_{[\sigma_c \to \sigma_c-1]}^-
\Phi(\vec{\gamma};\vec{\sigma};\vec{x}) 
\;.
\label{do2}
\end{eqnarray}
The linear differential operators $U_{\gamma_c \to \gamma_c+1}^+$, $L_{\sigma_c \to \sigma_c-1}^-$ are 
called {\it step-up} and {\it step-down} operators for the
upper and lower index, respectively. 
If additional step-down and step-up operators 
$U_{\gamma_c}^-$, $L_{\sigma_c}^+$ satisfying 
$$
U_{[\gamma_{c}+1 \to \gamma_c ]}^- U_{[\gamma_c \to \gamma_c+1]}^+ \Phi(\vec{\gamma},\vec{\sigma},\vec{x}) = 
L_{[\sigma_{c}-1 \to \sigma_{c}]}^+ L_{[\sigma_c \to \sigma_c-1]}^- \Phi(\vec{\gamma},\vec{\sigma},\vec{x}) =  
\Phi (\vec{\gamma},\vec{\sigma},\vec{x}) 
\;,
$$
({\it i.e.}, the inverses of  $U_{\gamma_c}^+$, $L_{\sigma_c}^-$), 
are constructed, we can combine these operators to shift the parameters of the 
hypergeometric function by any integer.
This process of applying $U_{\gamma_c}^\pm, L_{\sigma_c}^\pm$ to shift the
parameters by an integer is called 
{\bf differential reduction} of a hypergeometric function (\ref{Phi}).

Algebraic relations between functions 
$\Phi(\vec{\gamma},\vec{\sigma};\vec{x})$ 
with parameters shifted by integers are called {\bf contiguous relations}. 
The development of systematic techniques for obtaining a complete set of 
contiguous relations has a long story. 
It was started by Gauss, who found the differential reduction 
for the $_2F_1$ hypergeometric function in 1823. \cite{Gauss}
Numerous papers have since been published \cite{contiguous} on this problem.
An algorithmic solution was found by Takayama in Ref.\ \cite{theorem}, 
and those methods have been extended in a later 
\footnote{The problem also can be solved 
via an Ore algebra \cite{Ore} approach to the relevant 
system of linear differential and  difference (shift) operators.
}
series of publications \cite{Japan}. 

Let us recall that any hypergeometric function 
can be considered to be the solution of a proper system of partial 
differential equations (PDEs).  In particular, for a Horn-type hypergeometric 
function, the system of PDEs can be derived from the coefficients of the series.
In this case, the ratio of two coefficients can be presented as a ratio of 
two polynomials, 
\begin{equation}
\frac{C(\vec{m}+e_j)}{C(\vec{m})}  =  \frac{P_j(\vec{m})}{Q_j(\vec{m})} \;, 
\label{pre-diff}
\end{equation}
so that the Horn-type hypergeometric function satisfies the following system 
of equations:
\begin{equation}
0 = 
D_j (\vec{\gamma},\vec{\sigma},\vec{x})
\Phi(\vec{\gamma},\vec{\sigma}, \vec{x})
= 
\left[ 
Q_j\left( 
\sum_{k=1}^r x_k\frac{\partial}{\partial x_k}
\right)
\frac{1}{x_j} 
-
 P_j\left( 
\sum_{k=1}^r x_k\frac{\partial}{\partial x_k}
\right)
\right]
\Phi(\vec{\gamma},\vec{\sigma}, \vec{x}) \;, \quad j=1, \cdots, r.
\label{diff}
\end{equation}

Let $\mathfrak{R}$ be the left ideal of the ring $\mathfrak{D}$ of differential 
operators generated by the system of differential equations for a 
hypergeometric function (\ref{diff}), 
$D_j (\vec{\gamma},\vec{\sigma},\vec{x}),j=1,.\cdots,r$.
The first step in Takayama's algorithm
is the construction of a Gr\"obner basis 
$\mathfrak{G}=\left\{G_i(\vec{\gamma}, \vec{\sigma},\vec{x}), i=1,\cdots,q \right\}$ of  $\mathfrak{R}$.
Then the step-up operator corresponding to 
$
U_{\gamma_c}^+ 
$
and step-down operator corresponding to 
$
L_{\sigma_c}^-
$
are solutions to the linear equations  
\begin{eqnarray}
&& 
\sum_{i=1}^q C_i G_i(\vec{\gamma}, \vec{\sigma},\vec{x}) 
+ 
U_{[\gamma_{c}+1 \to \gamma_c ]}^- U_{[\gamma_c \to \gamma_c+1]}^+  = 1 \;, 
\\ && 
\sum_{i=1}^q E_i G_i(\vec{\gamma}, \vec{\sigma},\vec{x}) 
+ 
L_{[\sigma_{c}-1 \to \sigma_{c}]}^+ L_{[\sigma_c \to \sigma_c-1]}^- = 1\;,  
\end{eqnarray}
where 
$
C_i, E_i
$
are arbitrary functions. 
This system has a solution if the left ideal generated by 
$\mathfrak{G} \cup \left\{ U_{\gamma_c}^+ \right\}$
is equal to $\mathfrak{D}$ (see details in Ref.\ \cite{theorem}). 

In this way, the Horn-type structure provides an opportunity to reduce 
hypergeometric functions to a set of basis functions with 
parameters differing from the original values by integer shifts: 
\begin{equation}
P_0(\vec{x}) 
\Phi(\vec{\gamma}+\vec{k};\vec{\sigma}+\vec{l};\vec{x}) 
= 
\sum_{m_1, \cdots, m_p=0}^{\sum{|k_i|+\sum|l_i|}} P_{m_1, \cdots, m_r} (\vec{x}) 
\left( \frac{\partial}{\partial x_1} \right)^{m_1} \cdots
\left( \frac{\partial}{\partial x_r} \right)^{m_r} 
\Phi(\vec{\gamma};\vec{\sigma};\vec{x}) \;, 
\label{reduction}
\end{equation}
where $P_0(\vec{x})$ and $P_{m_1, \cdots, m_p}(\vec{x})$ are polynomials with 
respect to $\vec{\gamma},\vec{\sigma}$ and $\vec{x}$ and $\vec{k},\vec{l}$ 
are lists of integers. 

\section{Differential Reduction in Practice}
In real physical problems, the variables are generally not linearly 
independent: some of them can be equal to one another $x_i=x_j, i \neq j$. 
In this case, the Horn-type hypergeometric function generally is not 
expressible in terms of Horn hypergeometric functions with fewer variables.  
Another important physical case is when variables 
belong to the surfaces where coefficient $P_0(\vec{x})$ 
of Eq.\ (\ref{reduction}) 
vanishes (in the one-variable case it corresponds to $z=1$).
In all of these cases, the differential reduction cannot be directly applied.
But if the l.h.s.\ of Eq.\ (\ref{reduction}) is defined for that limit, 
the smooth limit of the r.h.s.\ of Eq.\ (\ref{reduction}) will exist too.

The problem is then to find this smooth limit.
Here, ``physics'' plays a role. For the evaluation of physical processes, 
exact results in terms of hypergeometric function are not necessary, 
but only the coefficients of a Laurent expansion around an integer value 
of the space-time dimension. 
From that point of view, to guarantee that smooth limit exists, 
it is enough to prove that a smooth limit exists for all coefficients of the 
all-order  $\ep$ expansion.

Recently, physicists have proven several theorems on the all-order 
$\ep$ expansion of hypergeometric functions about integer and/or 
rational values of parameters \cite{DK,nested,Kalmykov,KK08}.
A remarkable property of this construction is that for special values of 
parameters, the coefficients are expressible in terms of multiple 
polylogarithms \cite{mp}.
For functions of this type, the limiting procedure is well understood. 
Unfortunately, existing theorems are not adequate to cover all values of 
parameters for hypergeometric functions generated by Feynman 
diagrams \cite{KK08,sunset}.

\section{Discussion and Conclusions}
We have presented formal arguments that any Feynman diagram can be treated as 
a finite sum of Horn-type hypergeometric functions. 
This applies, in particular,
for off-shell diagrams with different values of masses in internal lines. 
In the physically interesting cases when some of the arguments
are equal to one another, or belong to some singularity surface, 
a limiting procedure should be constructed. 
To find this limit, and strongly prove that the proper limit exists, 
the all-order $\ep$ expansion of hypergeometric function around rational 
values of parameters can be used. 

The Horn-type hypergeometric functions possess useful properties: the
system of differential equation they satisfy is enough (i) for reduction 
of original function to a restricted set of basis functions (the number of 
basis functions follows directly from the system of equations); (ii) for the
construction of the all-order $\ep$ expansion for the basis hypergeometric 
functions in form of the Lappo-Danilevky solution.  The first part of this 
algorithm has been discussed in \cite{MKL06,LCWS08}; the second part in 
\cite{Kalmykov,KK08}.
The validity of our approach has been confirmed (in particular cases) 
by full agreement with the evaluation of the first coefficients of the 
$\ep$ expansion constructed in \cite{kalmykov:expansion}, by theorems 
about the all-order $\ep$ expansion of hypergeometric functions proven in 
\cite{nested} with the help of another technique, and by
comparisons of the results of the differential reduction of some Feynman 
diagrams with the results of reductions obtained via computer 
programs \cite{kalmykov:programs}. 
 
The above-mentioned properties of the hypergeometric representation, in 
particular Takayama's reduction algorithm, demonstrate that the hypergeometric 
representation is a universal tool for the evaluation of Feynman diagrams. 
Obtaining a decomposition of the integration 
region into regions where each individual term is well-defined is one of the 
main problems in the hypergeometric approach (beyond the one-loop and 
one-fold cases). 

Series representations (in four dimensions) have been used 
in Ref.\ \cite{Kershaw} to obtain a system of partial differential 
equations for $N$-point one-loop diagrams. In Ref.\ \cite{Davydychev}, the 
hypergeometric representation has been derived for an $N$-point one-loop 
diagram with arbitrary powers of the propagators. The possibility of using a
hypergeometric representation for the reduction of Feynman diagrams was 
considered in Ref.\ \cite{MKL06}.  The idea of using the Gr\"obner basis 
technique for the reduction of Feynman diagrams has been proposed by 
Tarasov \cite{Tarasov:Grobner} and received further extension 
in Ref.\ \cite{Grobner:difference}.
For practical applications of the ``limiting'' procedure, the first few 
coefficients of the $\ep$ expansion are necessary.  
The $\ep$ expansion is implemented in several packages \cite{packages} for
a restricted class of hypergeometric functions.  
Some results for the finite harmonic sums are available in \cite{finite}.

Another important class of developments includes techniques such as 
integration-by-parts \cite{ibp}, generalized recurrence 
relations \cite{numerator:massive}, 
and the differential equation approach \cite{DE}, which 
make it possible to work directly with 
parameters of the Feynman diagram (the l.h.s.\ of Eq.\ (\ref{FI})) without 
splitting the diagram into a linear combination of Horn-type hypergeometric 
functions.  Such direct analyses of Feynman diagrams as hypergeometric 
functions are based on 
the properties of dimensional regularization \cite{dimreg} , which
treats all types of singularities (IR and UV) simultaneously.

Some partial results of our research have been used in 
Ref.\ \cite{application}.  A more detailed 
discussion will be presented in a forthcoming publication \cite{BKK}.

\end{document}